\documentclass[prl,twocolumn,amsmath,amssymb,amsfonts,footinbib]{revtex4-2}

\usepackage{epsfig}
\usepackage{graphicx}
\usepackage{grffile} 
\usepackage{dcolumn}
\usepackage{bm}
\usepackage[titletoc]{appendix}

\usepackage{graphicx}
\usepackage{subfigure}
\usepackage{amsfonts}
\usepackage{amsgen}
\usepackage{amsbsy}
\usepackage{amsmath}
\usepackage{amssymb}
\usepackage{mathrsfs}
\usepackage{mathtools}
\usepackage{epstopdf}
\usepackage{caption}
\usepackage{tabularx}
\usepackage{tabu}
\usepackage{xcolor}
\usepackage{nicefrac}
\usepackage{enumitem}
\usepackage{gensymb}
\usepackage{natbib}
\usepackage{upgreek}
\usepackage{textcomp}  

\newcommand\etc{\textit{etc.\ }}
\newcommand\ie{\textit{i.e.~}}

\begin{document}

\title{Weakly adhesive suspension shows rate-dependence in oscillatory \\but not steady shear flows}

\author{Zhouyang Ge$^1$}\email{zhoge@mech.kth.se}
\author{Raffaella Martone$^2$}
\author{Luca Brandt$^1$}
\author{Mario Minale$^2$}\email{mario.minale@unicampania.it}

\affiliation{$^1$ Linn\'{e} FLOW Centre and Swedish e-Science Research Centre, 
Department of Engineering Mechanics, 
KTH Royal Institute of Technology, SE-100 44 Stockholm, Sweden}

\affiliation{$^2$ University of Campania ``Luigi Vanvitelli'', 
Department of Engineering, 
Real Casa dell'Annunziata, via Roma 29-81031 Aversa (CE), Italy}

\begin{abstract}
We report rheological measurements of a noncolloidal particle suspension in a Newtonian solvent 
at 40\% solid volume fraction. 
An anomalous, frequency-dependent complex viscosity is found under oscillatory shear (OS) flow, 
whereas a constant dynamic viscosity is found under the same shear rates in steady shear (SS) flow. 
We show that this contradiction arises from the underlying microstructural difference between OS and SS, 
mediated by weak interparticle forces. 
Discrete element simulations of proxy particle suspensions confirm this hypothesis and 
reveal an adhesion-induced, shear thinning mechanism with a $-1/5$ slope, only in OS, in agreement with experiments.
\end{abstract}

\maketitle


Dense particle suspensions in a Newtonian fluid are known to display a rich variety of rheologies,
such as shear thickening/thinning, yielding, ageing, \etc \cite{mewis_wagner_book,guazzelli_pouliquen_2018, 
Morris_annurev-fluid, shear_thinning_SM18}.
These complex rheological behaviours are strongly affected by the details of the underlying microstructure,
which, apart from being a direct consequence of the sample preparation,
is intimately related to various hydrodynamic/interparticle interactions,
including e.g.\ lubrication, Brownian, electrostatic, van der Waals and contact forces \cite{Mewis_Wagner_2009}.
The presence of a non-hydrodynamic force, $\mathcal{F}$, 
introduces a second time scale, $\uptau_\mathcal{F}\equiv 6\pi\eta_f a^2/\mathcal{F}$
($\eta_f$ denotes the fluid dynamic viscosity and $a$ the characteristic particle radius),
in addition to that imposed by the shear, $\uptau_s\equiv \dot{\gamma}^{-1}$
($\dot{\gamma}$ the shear rate).
And, if $\uptau_\mathcal{F}/\uptau_s$ is less than a critical threshold
\footnote{The threshold value of $\uptau_\mathcal{F}/\uptau_s$, 
above which the suspension rheology becomes hydrodynamically dominated,
depends on the particle concentration. 
For dense suspensions, this value can be much larger than unity 
as the characteristic hydrodynamic force is $\gg 6\pi\eta_f a^2\dot{\gamma}$.},
a nonlinear rate-dependent rheology may be expected when increasing $\uptau_\mathcal{F}/\uptau_s$,
as the suspension transits from interactions dominated by non-hydrodynamic to hydrodynamic forces, 
cf.\ \cite{hinch_2011,Seto_PRL2013,Mari_2014JOR}.

While generally true for dense, overdamped suspensions in the Stokes flow regime, 
the above simple analysis makes no distinction between steady shear (SS) and oscillatory shear (OS) flows.
Certainly, the Cox-Merz rule is not universally valid
\footnote{The Cox-Merz rule states that the dynamic viscosity of a fluid in SS is 
equal to its complex viscosity in OS at matching frequency, 
\ie $\eta(\dot{\gamma})=\eta^*(\omega)|_{\omega=\dot{\gamma}}$; 
see Ref.\ \cite{mewis_wagner_book}.},
as the suspension microstructure is strain-dependent in OS \cite{Bricker_Butler2006,Bricker_Butler2007};
the next question is whether the same \emph{rate}-dependence hold for both SS and OS 
within the same range of $\uptau_\mathcal{F}/\uptau_s$.
In the literature, it is generally assumed that if SS is rate-independent, OS is also frequency-independent.
However, recent experiments challenge this assumption and
demonstrate a frequency-dependent rheology in the \emph{absence} of any rate-dependence in SS
\cite{Carotenuto_conf_2014,Martone_conf_2018,Martone_experiment}.
In this Letter, we combine experiments with simulations and 
show that \emph{weak} repulsive or adhesive interparticle forces 
are enough to induce the sought frequency-dependence in OS, while keeping SS rate-independent.
Currently, there is a growing interest in elucidating the rheology of dense particle suspensions,
motivated by both fundamental quests in nonequilibrium statistical mechanics 
and engineering applications, cf.\ \cite{dense_preface2020}.
We believe the present finding is an important step toward this goal
and may guide judicious designs of complex fluids and functional metamaterials.

\emph{Experiments.---}  
Time sweep oscillatory tests are performed on a suspension of glass hollow microspheres 
dispersed in a Newtonian fluid (polyisobutene) at volume fraction $\phi=40\%$.
The experiments are executed on a constant-strain rheometer, 
ARES G2 (TA Instruments) equipped with a cone-and-plate geometry, 
by imposing a sinusoidal strain during each run. 
The amplitude of the applied strain, $\gamma_0$, is varied from 0.5\% to 200\%. 
Three angular frequencies, $\omega$ = 5, 50 and 200 rad/s, are tested repeatedly.
After a steady preconditioning shear, the complex viscosity is followed in time, 
or equivalently, as suggested in \cite{Bricker_Butler2006}, vs.\ the total accumulated strain, 
$\gamma_{tot} = 4 \gamma_0 n_{cyc}$, where $n_{cyc}$ is the number of cycles of the oscillatory shear.
Further details on the experiments are available in \emph{Supplementary Information} (SI) and \cite{Martone_experiment}.

\begin{figure*}[t]
 \begin{center}
 \includegraphics[width=\textwidth]{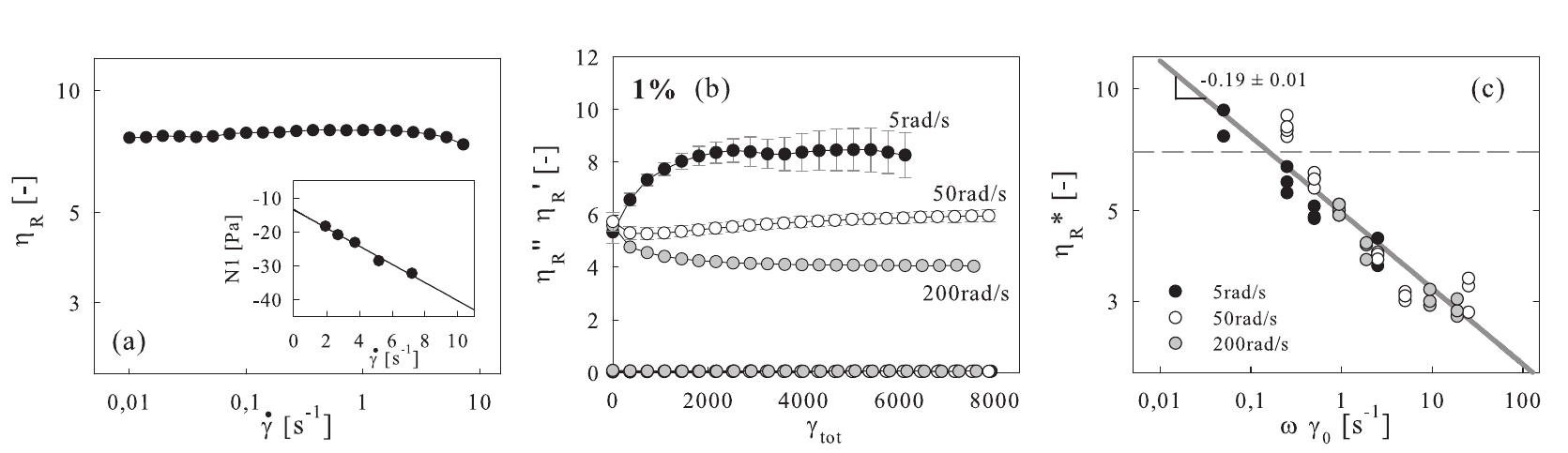}
 \end{center}
 \caption{Experimental results. 
   (a) Steady flow curves of the suspension: relative viscosity ($\eta_R$) and first normal stress difference ($N_1$, inset) vs.\ shear rate.
   (b) Evolution of the dynamic viscosity ($\eta'_R$) and its elastic counterpart ($\eta''_R$) vs.\ the total accumulated strain at $\gamma_0=1\%$. The angular frequencies are indicated next to $\eta'_R$.
   (c) Relative complex viscosity ($\eta^*_R$) as a function of the maximal shear rate. The solid line is a best fit with slope $-0.19 \pm 0.01$. The dashed line corresponds to the averaged value of $\eta_R$ in (a).}
 \label{fig:exp raw}
\end{figure*}

The suspension is interialess and non-Brownian, 
as the particle Reynolds number is smaller than $10^{-6}$, 
and the P\'eclet number (Pe) is larger than $10^{5}$. 
Particle sedimentation can be neglected, as the average Shields number is about $10^3$. 
A characteristic time arising from Brownian diffusion or sedimentation is thus irrelevant in the investigated suspension 
and, accordingly, the SS behaviour shows constant viscosity 
and first normal stress difference negative and linear in the shear rate; 
see Fig.\ \ref{fig:exp raw}(a). 
In OS, the complex viscosity, $\eta^* \equiv \eta'-i\eta''$, 
evolves in time and is a function of $\gamma_0$, 
as widely reported in the literature \cite{Bricker_Butler2006, Breedveld_etal_2001, Corte_NatPhys_2008};
but, surprisingly, it also depends on $\omega$.
In Fig.\ \ref{fig:exp raw}(b), we plot the relative dynamic viscosity, $\eta'_R$, 
and the out-of-phase component, $\eta''_R$, of the relative complex viscosity vs.\ $\gamma_{tot}$, 
for $\gamma_0=1$\% and three different $\omega$. 
$\eta'_R$ is always about two orders-of-magnitude larger than $\eta''_R$, 
thus it is practically coincident with $\eta^{*}_R$, 
highlighting the viscous behaviour of the suspension. 
Fig.\ \ref{fig:exp raw}(b) also shows the viscosity $\eta'_R$ varies with $\omega$.
In general, we identify an $\omega$-dependent regime for $\gamma_0<1$ and 
an $\omega$-independent regime for larger $\gamma_0$ \cite{Martone_experiment}. 
These two regimes coincide with those observed by Lin et al.\ \cite{Lin_Phan-Thien_Khoo_2013}, 
who showed that in the first regime the microstructure self-arrangement is driven by the shear-induced particle diffusion, 
while in the second one the microstructure is immediately formed by the oscillation itself, 
similarly to what happens for a steady flow reversal that is indeed rate-independent. 
Moreover, in the first regime, the plateau values reached at large $\gamma_{tot}$ of data 
taken at different $\omega$ and $\gamma_0$ collapse on a single master curve
if $\eta^{*}_R$ is plotted vs.\ the maximum shear rate $\omega\gamma_0$ 
(Fig.\ \ref{fig:exp raw}c).
That is, the suspension is 
\emph{shear thinning} with respect to an increasing shear rate over three decades, 
and the power law exponent is about $-1/5$.
This rate-dependence suggests there must be a non-hydrodynamic force in the suspension, 
which is only relevant at $\gamma_0 <1$.
In the following, we trace the origin of this rheology using numerical simulations.

\emph{Simulations.---}
To investigate the anomalous frequency dependence, 
we use a minimal numerical model based on the hybrid lubrication/granular dynamics
\cite{Mari_2014JOR, Cheal_Ness_2018, hlgd}. 
In this framework, particle dynamics are obtained by 
approximating the hydrodynamics with near-neighbour lubrication force, 
combined with interparticle physiochemical or contact interactions.
Below, we present simulation results of 500 spherical particles in a cube ($\phi=40\%$) 
subject to the Lees-Edwards boundary condition, sheared either steadily or periodically.
Both monodisperse and bidisperse (radius ratio 1.4) suspensions are considered to check the size effect, 
for which we observe no major difference.
The suspension responds in the non-inertial limit,
as we ensure the sum of all interactions is much less than the reference force scales.
See SI for the complete numerical details and visualizations.

\begin{figure*}[t]
 \begin{center}
 \includegraphics[width=0.9\textwidth]{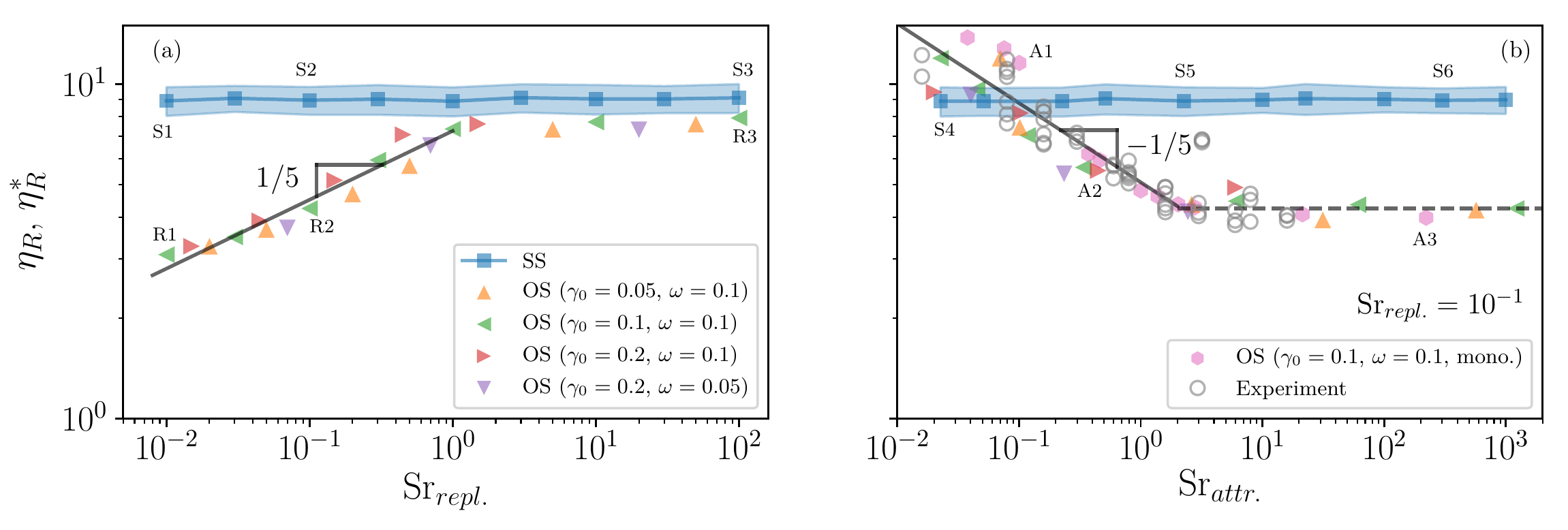}
 \begin{picture}(0,0)
   \setlength{\unitlength}{\columnwidth}
   \put(-1.63,0.14){\includegraphics[height=1.55cm]{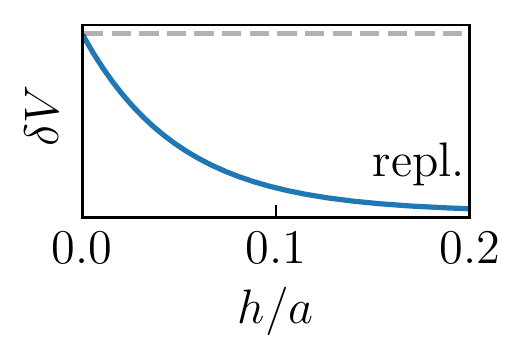}}
   \put(-0.8, 0.14){\includegraphics[height=1.7 cm]{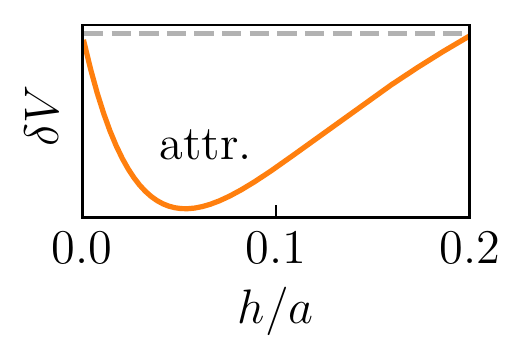}}
 \end{picture}
 \end{center}
 \caption{Simulation results of (a) repulsive and (b) adhesive suspensions.
   In both cases, $\eta_R$ is nearly constant under SS (shades denote one standard deviation);
   whereas $\eta^*_R$ displays either shear thickening or thinning in OS 
   (slopes and numbers are least squares fits and power law exponents).
   Insets show generic interaction potentials as functions of surface gap (dashed lines denote $\delta V=0$).
   The experimental data is reproduced in (b).}
 \label{fig:visc}
\end{figure*}

\emph{Discussions.---}
To begin with, we notice from the literature that shear thinning of dense, non-Brownian suspensions 
has been attributed to several mechanisms
\footnote{There is yet another phenomenological mechanism for shear thinning -- stress relaxation.
Since the sample is presheared in SS then measured in OS, at $\gamma_0 \lessapprox 1$, 
the previously accumulated stress may be released to a greater extent in lower $\gamma_0$.
However, such relaxation has no time dependence (no additional force scale), thus cannot explain the present experiment.}:
non-Newtonian properties of the solvent \cite{NN_solvent0,NN_solvent1},
particle softness \cite{Kalman_Rosen_Wagner2008, ranga_thinning},
inertia \cite{Irani_prf2019},
variable friction coefficient \cite{lobry_lemaire_blanc_gallier_peters_2019},
interparticle repulsion \cite{Mari_2014JOR, shear_thinning_SM18} as well as
attraction \cite{Brown_nmat2010,Singh_attr_prl2019}.

The first three possibilities are immediately eliminated, 
since we work with a Newtonian solvent and hard spheres in the overdamped regime;
the external stress is simply too low for non-Newtonian behaviours or particle deformations to be expected 
\footnote{The maximal shear stress in the experiment is $\sigma_s=1120$ Pa, 
way below the elastic modulus of the glass microspheres ($G \approx 70$ GPa). 
Thus, the deformation is negligible ($\sigma_s/G < 1.6\times 10^{-8}$).}.

As for friction, our simulations show negligible particle contact in oscillatory flow at $\phi=40\%$;
the average particle overlap is typically $\sim \mathcal{O}(10^{-5}a)$ (see SI).
Therefore, friction is seldom active in OS.
Furthermore, even if the friction coefficient reduces with the normal load 
as in the model in \cite{lobry_lemaire_blanc_gallier_peters_2019},
it should have stronger effects in SS.
We have checked that including the contact model of \cite{lobry_lemaire_blanc_gallier_peters_2019} in our system
leads to shear-thinning in SS rather than OS, 
contrary to what the experiments show.
Therefore, an explanation in terms of a variable friction coefficient is also unlikely.

\begin{figure}[t]
  \begin{center}
    \includegraphics[width=0.9\columnwidth]{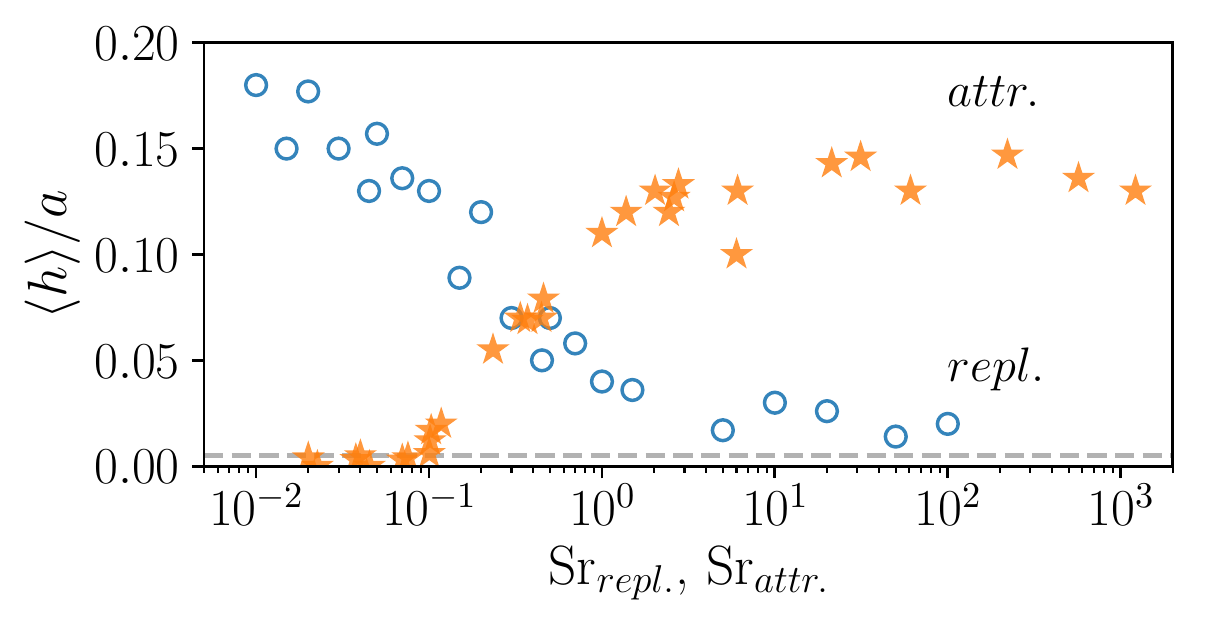}
  \end{center}
  \caption{Medians of the average surface gap.
  Circles and stars denote the repulsive and adhesive suspensions in OS, respectively.
  The dashed line corresponds to $\langle h \rangle_{SS}$.}
  \label{fig:median}
\end{figure}

To examine the effect of interparticle forces,
we impose pairwise ($ij$) electrostatic repulsion and van der Waals attraction,
given as \cite{mewis_wagner_book,Israelachvili_book}
\begin{equation} \label{force-balance}
  {\bm F}_{repl.}= - \mathcal{F}_r \frac{e^{-h\kappa}}{a/\bar{a}} {\bm n}, \quad
  {\bm F}_{attr.}= \frac{A \bar{a} {\bm n}}{12(h^2+\epsilon^2)} ,
\end{equation}
where $\mathcal{F}_r$ is a repulsive force scale,
$\bm n$ the normal vector,
$\bar{a}=2a_ia_j/(a_i+a_j)$ the harmonic mean radius,
$\kappa^{-1}=0.05a$ the Debye length,
$A$ the Hamaker constant,
and $\epsilon=0.1\bar{a}$ a regularization term, cf.\ \cite{Mari_2014JOR,Singh_attr_prl2019}.
We activate ${\bm F}_{repl.}$ and/or ${\bm F}_{attr.}$ when $h \leq 0.2a$, 
and saturate them by setting $h = 0$ when overlap occurs.
The corresponding interaction potentials are illustrated in Fig.\ \ref{fig:visc} (insets),
where the adhesive potential is a superposition of both repulsion and attraction.
In the latter case, an adhesive force scale can be defined as 
$\mathcal{F}_a =A \bar{a}/(12\langle h \rangle^2 +12\epsilon^2)$,
where $\langle h \rangle$ is the median surface gap ($h$) sampled from all interacting pairs (see Fig.\ \ref{fig:median}).

As discussed earlier, the effect of a non-hydrodynamic force, $\mathcal{F}$, 
on the suspension rheology can be measured by the dimensionless group, 
$\uptau_\mathcal{F}/\uptau_s$.
Specifically, we write
\begin{equation} \label{eq:Sr}
  \textrm{Sr}_{repl.} \equiv \frac{b \gamma_0 \omega}{\mathcal{F}_r}, \quad
  \textrm{Sr}_{attr.} \equiv \frac{b \gamma_0 \omega}{\mathcal{F}_a},
\end{equation}
where $b=6\pi\eta_f a^2c$, with $c$ a $\phi$-dependent prefactor
\footnote{See Fig.\ 10 of Ref.\ \cite{Martone_experiment} for evidence of the $\phi$ dependence.}.
Here, Sr$_{repl.}$ and Sr$_{attr.}$ represent relative shear rates,
analogously to the Weissenberg number in steady flows 
(in which case $\gamma_0 \omega$ should be replaced by $\dot{\gamma}$) 
and the Deborah number in transient flows \cite{Poole_Wi_De,mewis_wagner_book}. 
The critical value, Sr$_{crit}$, above which the suspension is completely Newtonian, is not known \emph{a priori}.
Here, Sr$_{crit} \approx 1$ if we set  $c= 10^{-3}$.

Fig.\ \ref{fig:visc} shows the main result of the simulations.
One immediate observation is that the suspension remains rate-independent in SS regardless of any interparticle forces;
$\eta_R$ is nearly constant across four decades of Sr, 
consistent with Fig.\ \ref{fig:exp raw}(a) and extending its range of validity.
In stark contrast, $\eta_R^*$  exhibits either shear thickening or thinning in OS,
depending on the direction of the force.
With repulsive potentials, $\eta_R^* \propto$ Sr$_{repl.}^{1/5}$;
while in the adhesive case, $\eta_R^* \propto$ Sr$_{attr.}^{-1/5}$.
The trends hold for all strain amplitudes ($\gamma_0$), angular frequencies ($\omega$) and size ratios considered.
In particular, the agreement between the fitted slopes from adhesive suspensions and experiments implies that 
it is the interparticle attraction that produces the additional frequency dependence in OS. 
Indeed, as we shift the experimental data vertically to match the corresponding $\eta_R$ in SS 
and horizontally to match Sr$_{crit.}$, 
the two data sets collapse around a single line (see Fig.\ref{fig:visc}b).
Given that the exact magnitude of the adhesive forces is unknown
(indicating that the slope $-1/5$ is not sensitive to the details of the force),
and $\mathcal{F}_a \ll 6\pi\eta_f a^2\gamma_0\omega$ even at its peak (see SI),
this is a rather remarkable result.

In the remainder, we explore how such rate-dependent rheologies arise from interparticle forces only in OS but not SS.
As we have introduced additional forces in the suspension, 
a natural question is how much they contribute to the final stress.
This can be readily extracted from our simulations, 
where the bulk stress tensor is computed by summing the various one-body or pairwise stresslets 
due to the Stokes drag, lubrication and non-hydrodynamic forces,
$\Sigma_{mn}= 2\eta_f{E}_{mn}^\infty +
\langle { S_{mn}^S} \rangle_{i}+ 
\langle { S_{mn}^L} \rangle_{ij} +
\sum \langle { S_{mn}^F} \rangle_{ij}$, 
where $E_{mn}^\infty$ is the rate-of-strain tensor. 
These are averaged over either all particles or interacting pairs \cite{hlgd}.

\begin{figure}[b]
 \begin{center}
 \includegraphics[width=\columnwidth]{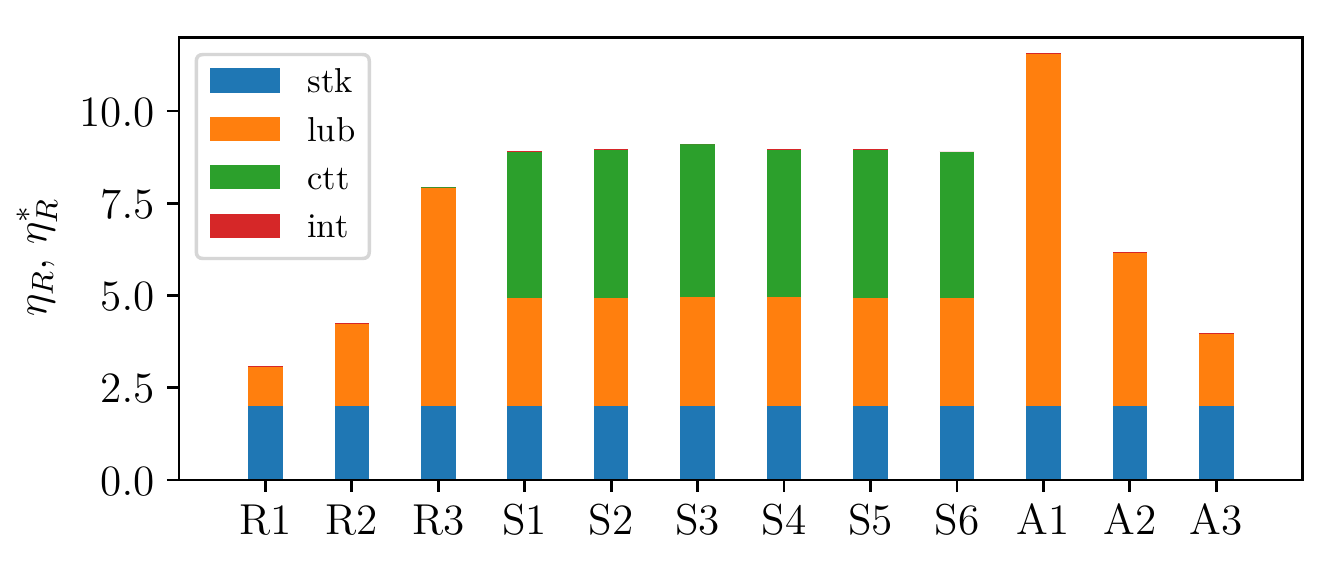}
 \end{center}
 \caption{Shear stress components of representative SS and OS cases under repulsive and adhesive interactions; 
   see Fig.\ \ref{fig:visc} for the annotation. 
 }
 \label{fig:budget}
\end{figure}

Fig.\ \ref{fig:budget} illustrates the shear stress ($xy$) decomposition.
First, we notice that the Stokes drag contributes identically in all cases,
which is trivial as $S_{xy}^S$ only depends on $\phi$ by definition \cite{hlgd}.
Second, the lubrication and contact stresses are nearly constant in SS,
whereas in OS, the former is shear-dependent and the latter is vanishing.
The absence of particle contact in OS suggests that 
the microstructure organizes into \emph{absorbing states} \cite{Corte_NatPhys_2008};
particles follow nonaffine trajectories in the gap of each other during each oscillation cycle, 
but always return to isotropic arrangements at zero amplitudes. 
Mathematically, this corresponds to $\langle {\bm r}{\bm r} \rangle_{xy}=0$ at $\gamma=0$, 
where $\bm r$ is the center-to-center position vector between two neighbouring particles \cite{Ness_OS_2017SM}.
We have checked that the above condition is met in all OS cases (see SI).
The result of zero contact stress is thus self-consistent.

\begin{figure}[t]
  \begin{center}
    \includegraphics[width=\columnwidth]{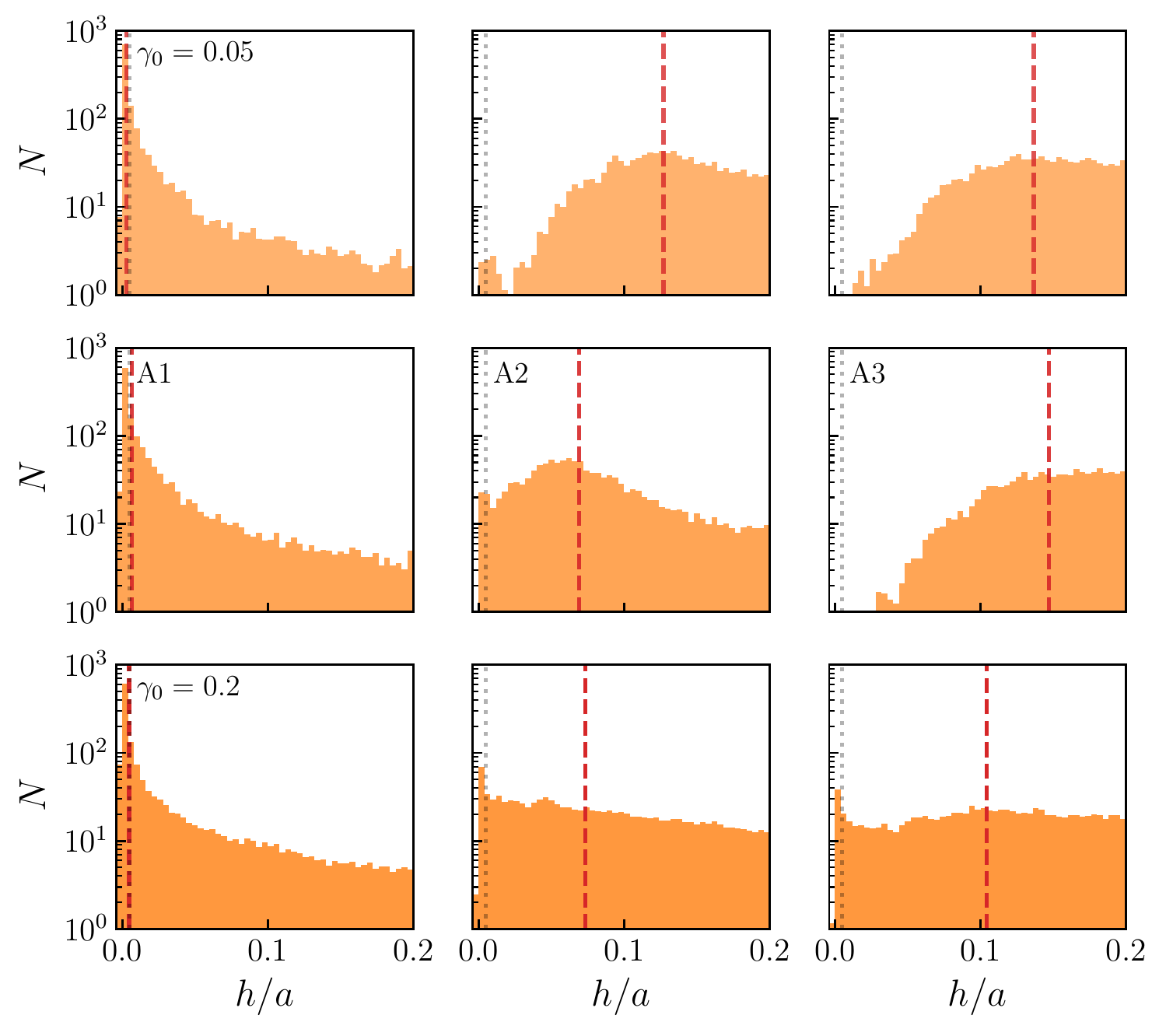}
    \begin{picture}(0,0)
      \setlength{\unitlength}{\columnwidth}
      \put(-0.3,0.47){\includegraphics[height=1.5cm]{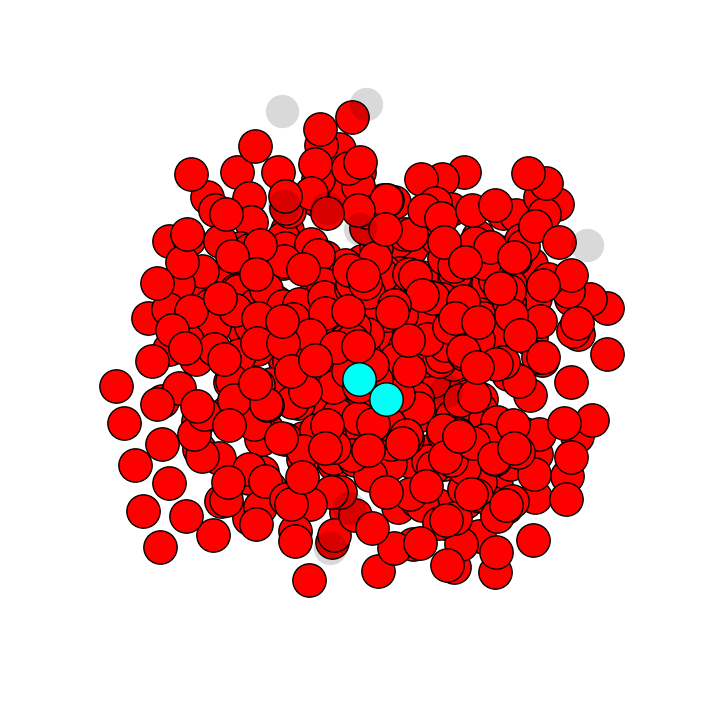}}
      \put(0.01,0.47){\includegraphics[height=1.5cm]{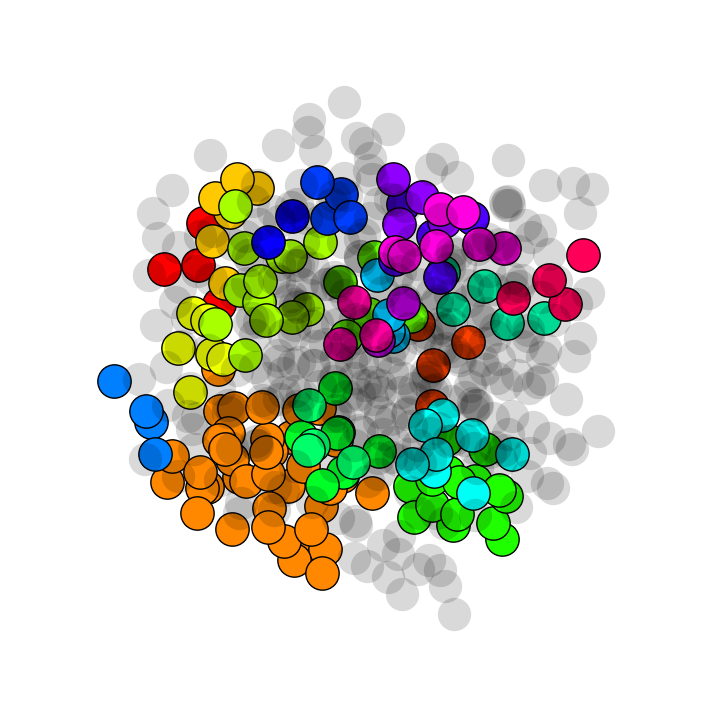}}
      \put(0.2 ,0.47){\includegraphics[height=1.5cm]{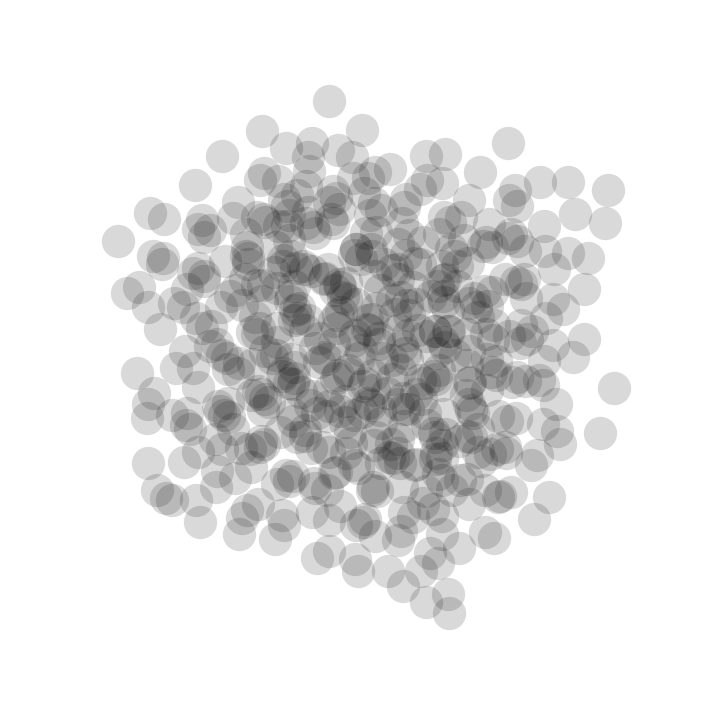}}
    \end{picture}
  \end{center}
  \caption{Surface gap distributions under increasing Sr$_{attr.}$ (left to right) 
    and $\gamma_0$ (top to bottom) in OS, averaged over the last ten cycles.
    $N$ is the number of pairs.
    Dashed lines (red) denote $\langle h \rangle$. 
    Dotted lines (gray) denote the average $\langle h \rangle_{SS}$, obtained from cases (S1, S2, S3). 
    Insets in the mid-panel are snapshots of the hydroclusters in the steady-states.}
  \label{fig:hist}
\end{figure}

The third and perhaps most unexpected finding is that stresses due to interparticle repulsions or attractions 
are completely negligible in all cases (Fig.\ \ref{fig:budget}, red).
Since the rate-dependent relative viscosity results from such forces,
naively, one would expect they also make a rate-dependent contribution to the stress budget, 
or at least be active in OS. 
In contrast, we see nearly zero $\mathcal{F}-$components across four decades of shear rate, 
regardless of SS or OS. 
This can only be explained by its small magnitude.
With $\mathcal{F}_a =A \bar{a}/(12\langle h \rangle^2 +12\epsilon^2) \sim A/\langle h \rangle^2$,
the Peclet number based on $\mathcal{F}_a$ will be in the range $\mathcal{O}(10^1 \sim 10^3)$ (see SI).
Hence, Pe $\gg 1$ does not necessarily guarantee a negligible ``thermal'' effect.
Here, the weak adhesive (or repulsive) force clearly represents a \emph{nudge} 
for the evolution of the suspension microstructure,
evidenced by the lubrication stress,
which ultimately alters $\eta_R^*$
\footnote{An alternative measure of the microstructure evolution is particle diffusivities, cf.\ \cite{Pine_Nature_2005}.
We have checked the cases of $\gamma_0=0.1$ and $\omega=0.1$ under different Sr$_{attr}$, 
and found negligible self-diffusivities throughout.}.

In the above, we use $\langle h \rangle$ to quantify the role of the adhesive force,
as the \emph{effective} interaction is a statistical quantity.
Fig.\ \ref{fig:hist} shows the average gap distributions for nine different adhesive suspensions 
under different Sr$_{attr.}$ and $\gamma_0$.
The left-most bins, indicating the number of pairs in contact, 
are only occupied at lower shear conditions.
As the relative shear increases, \ie the attractive force reduces, 
$\langle h \rangle$ shifts to higher values and the effective attraction further decreases. 
This reinforces the shear thinning and is observed regardless of $\gamma_0$.
On the other hand, repulsive suspensions experience the opposite
with the effective repulsion increasing when increasing the imposed shear (see Fig.\ \ref{fig:median} and SI).
Finally, the median gap in SS, $\langle h \rangle_{SS}$, is insensitive to shear as expected from its constant viscosity.

Complementary to $\langle h \rangle$, we also show flow-induced hydroclusters (HC) 
for cases (A1, A2, A3) in Fig.\ \ref{fig:hist}.
Here, particles
are depicted with the same color if adjacent,
with ``noise'' particles in gray (see SI for details).
Clearly, the size of HC reaches the length scale of the domain at the lowest shear;
as the shear increases, it splits into smaller satellite HC and eventually all disappear.
This provides another view for the shear thinning rheology --
as the number density is lower outside HC, the fluid is under less stress,
which leads to a decrease in energy dissipation and thus a reduced viscosity \cite{Wagner_Brady_2009}.

\emph{Summary.---}
Experiments of a noncolloidal particle suspension showing a frequency-dependent complex viscosity 
and a constant dynamic viscosity are discussed.
Using minimal numerical simulations, 
we identify an adhesion-induced, shear thinning mechanism that is 
insensitive to the specifics of the weak force and generic to time-periodic flows,
thus explaining the experiments.
In general, our finding is not only relevant to rheological characterizations of suspension-based functional materials,
but may also inspire further theoretical development of self-organization and absorbing-state transitions 
in the presence of another actor ($\mathcal{F}$) \cite{Corte_etal_PRL2009, Royer49, Ness_Cates_PRL2020}.


\medskip

We thank R. Radhakrishnan for in-depth discussions.
We also thank C. Ness, F. Peters and A. Los for useful suggestions.
The work is supported by the Swedish Research Council (grant no.\ VR 2014--5001)
and University of Campania `L. Vanvitelli' under the  programme ``VALERE: VAnviteLli pEr la RicErca'' project: SEND.

\bibliographystyle{apsrev4-2}
\small
\bibliography{main}

\begin{thebibliography}{41}%
\makeatletter
\providecommand \@ifxundefined [1]{%
 \@ifx{#1\undefined}
}%
\providecommand \@ifnum [1]{%
 \ifnum #1\expandafter \@firstoftwo
 \else \expandafter \@secondoftwo
 \fi
}%
\providecommand \@ifx [1]{%
 \ifx #1\expandafter \@firstoftwo
 \else \expandafter \@secondoftwo
 \fi
}%
\providecommand \natexlab [1]{#1}%
\providecommand \enquote  [1]{``#1''}%
\providecommand \bibnamefont  [1]{#1}%
\providecommand \bibfnamefont [1]{#1}%
\providecommand \citenamefont [1]{#1}%
\providecommand \href@noop [0]{\@secondoftwo}%
\providecommand \href [0]{\begingroup \@sanitize@url \@href}%
\providecommand \@href[1]{\@@startlink{#1}\@@href}%
\providecommand \@@href[1]{\endgroup#1\@@endlink}%
\providecommand \@sanitize@url [0]{\catcode `\\12\catcode `\$12\catcode
  `\&12\catcode `\#12\catcode `\^12\catcode `\_12\catcode `\%12\relax}%
\providecommand \@@startlink[1]{}%
\providecommand \@@endlink[0]{}%
\providecommand \url  [0]{\begingroup\@sanitize@url \@url }%
\providecommand \@url [1]{\endgroup\@href {#1}{\urlprefix }}%
\providecommand \urlprefix  [0]{URL }%
\providecommand \Eprint [0]{\href }%
\providecommand \doibase [0]{https://doi.org/}%
\providecommand \selectlanguage [0]{\@gobble}%
\providecommand \bibinfo  [0]{\@secondoftwo}%
\providecommand \bibfield  [0]{\@secondoftwo}%
\providecommand \translation [1]{[#1]}%
\providecommand \BibitemOpen [0]{}%
\providecommand \bibitemStop [0]{}%
\providecommand \bibitemNoStop [0]{.\EOS\space}%
\providecommand \EOS [0]{\spacefactor3000\relax}%
\providecommand \BibitemShut  [1]{\csname bibitem#1\endcsname}%
\let\auto@bib@innerbib\@empty
\bibitem [{\citenamefont {Mewis}\ and\ \citenamefont
  {Wagner}(2012)}]{mewis_wagner_book}%
  \BibitemOpen
  \bibfield  {author} {\bibinfo {author} {\bibfnamefont {J.}~\bibnamefont
  {Mewis}}\ and\ \bibinfo {author} {\bibfnamefont {N.~J.}\ \bibnamefont
  {Wagner}},\ }\href@noop {} {\emph {\bibinfo {title} {Colloidal suspension
  rheology}}}\ (\bibinfo  {publisher} {{C}ambridge {U}niversity {P}ress},\
  \bibinfo {year} {2012})\BibitemShut {NoStop}%
\bibitem [{\citenamefont {Guazzelli}\ and\ \citenamefont
  {Pouliquen}(2018)}]{guazzelli_pouliquen_2018}%
  \BibitemOpen
  \bibfield  {author} {\bibinfo {author} {\bibfnamefont {E.}~\bibnamefont
  {Guazzelli}}\ and\ \bibinfo {author} {\bibfnamefont {O.}~\bibnamefont
  {Pouliquen}},\ }\href {https://doi.org/10.1017/jfm.2018.548} {\bibfield
  {journal} {\bibinfo  {journal} {J. Fluid Mech.}\ }\textbf {\bibinfo {volume}
  {852}},\ \bibinfo {pages} {P1} (\bibinfo {year} {2018})}\BibitemShut
  {NoStop}%
\bibitem [{\citenamefont {Morris}(2020)}]{Morris_annurev-fluid}%
  \BibitemOpen
  \bibfield  {author} {\bibinfo {author} {\bibfnamefont {J.~F.}\ \bibnamefont
  {Morris}},\ }\href {https://doi.org/10.1146/annurev-fluid-010816-060128}
  {\bibfield  {journal} {\bibinfo  {journal} {Annual Review of Fluid
  Mechanics}\ }\textbf {\bibinfo {volume} {52}},\ \bibinfo {pages} {121}
  (\bibinfo {year} {2020})},\ \Eprint
  {https://arxiv.org/abs/https://doi.org/10.1146/annurev-fluid-010816-060128}
  {https://doi.org/10.1146/annurev-fluid-010816-060128} \BibitemShut {NoStop}%
\bibitem [{\citenamefont {Chatt\'e}\ \emph {et~al.}(2018)\citenamefont
  {Chatt\'e}, \citenamefont {Comtet}, \citenamefont {Niguès}, \citenamefont
  {Bocquet}, \citenamefont {Siria}, \citenamefont {Ducouret}, \citenamefont
  {Lequeux}, \citenamefont {Lenoir}, \citenamefont {Ovarlez},\ and\
  \citenamefont {Colin}}]{shear_thinning_SM18}%
  \BibitemOpen
  \bibfield  {author} {\bibinfo {author} {\bibfnamefont {G.}~\bibnamefont
  {Chatt\'e}}, \bibinfo {author} {\bibfnamefont {J.}~\bibnamefont {Comtet}},
  \bibinfo {author} {\bibfnamefont {A.}~\bibnamefont {Niguès}}, \bibinfo
  {author} {\bibfnamefont {L.}~\bibnamefont {Bocquet}}, \bibinfo {author}
  {\bibfnamefont {A.}~\bibnamefont {Siria}}, \bibinfo {author} {\bibfnamefont
  {G.}~\bibnamefont {Ducouret}}, \bibinfo {author} {\bibfnamefont
  {F.}~\bibnamefont {Lequeux}}, \bibinfo {author} {\bibfnamefont
  {N.}~\bibnamefont {Lenoir}}, \bibinfo {author} {\bibfnamefont
  {G.}~\bibnamefont {Ovarlez}},\ and\ \bibinfo {author} {\bibfnamefont
  {A.}~\bibnamefont {Colin}},\ }\href {https://doi.org/10.1039/C7SM01963G}
  {\bibfield  {journal} {\bibinfo  {journal} {Soft Matter}\ }\textbf {\bibinfo
  {volume} {14}},\ \bibinfo {pages} {879} (\bibinfo {year} {2018})}\BibitemShut
  {NoStop}%
\bibitem [{\citenamefont {Mewis}\ and\ \citenamefont
  {Wagner}(2009)}]{Mewis_Wagner_2009}%
  \BibitemOpen
  \bibfield  {author} {\bibinfo {author} {\bibfnamefont {J.}~\bibnamefont
  {Mewis}}\ and\ \bibinfo {author} {\bibfnamefont {N.}~\bibnamefont {Wagner}},\
  }\href@noop {} {\bibfield  {journal} {\bibinfo  {journal} {J. Non-Newtonian
  Fluid Mech.}\ }\textbf {\bibinfo {volume} {157}},\ \bibinfo {pages} {147}
  (\bibinfo {year} {2009})}\BibitemShut {NoStop}%
\bibitem [{Note1()}]{Note1}%
  \BibitemOpen
  \bibinfo {note} {The threshold value of $\uptau _\protect \mathcal {F}/\uptau
  _s$, above which the suspension rheology becomes hydrodynamically dominated,
  depends on the particle concentration. For dense suspensions, this value can
  be much larger than unity as the characteristic hydrodynamic force is $\gg
  6\pi \eta _f a^2\protect \mathaccentV {dot}05F{\gamma }$.}\BibitemShut
  {Stop}%
\bibitem [{\citenamefont {Hinch}(2011)}]{hinch_2011}%
  \BibitemOpen
  \bibfield  {author} {\bibinfo {author} {\bibfnamefont {E.~J.}\ \bibnamefont
  {Hinch}},\ }\href {https://doi.org/10.1017/jfm.2011.350} {\bibfield
  {journal} {\bibinfo  {journal} {J. Fluid Mech.}\ }\textbf {\bibinfo {volume}
  {686}},\ \bibinfo {pages} {1–4} (\bibinfo {year} {2011})}\BibitemShut
  {NoStop}%
\bibitem [{\citenamefont {Seto}\ \emph {et~al.}(2013)\citenamefont {Seto},
  \citenamefont {Mari}, \citenamefont {Morris},\ and\ \citenamefont
  {Denn}}]{Seto_PRL2013}%
  \BibitemOpen
  \bibfield  {author} {\bibinfo {author} {\bibfnamefont {R.}~\bibnamefont
  {Seto}}, \bibinfo {author} {\bibfnamefont {R.}~\bibnamefont {Mari}}, \bibinfo
  {author} {\bibfnamefont {J.~F.}\ \bibnamefont {Morris}},\ and\ \bibinfo
  {author} {\bibfnamefont {M.~M.}\ \bibnamefont {Denn}},\ }\href
  {https://doi.org/10.1103/PhysRevLett.111.218301} {\bibfield  {journal}
  {\bibinfo  {journal} {Phys. Rev. Lett.}\ }\textbf {\bibinfo {volume} {111}},\
  \bibinfo {pages} {218301} (\bibinfo {year} {2013})}\BibitemShut {NoStop}%
\bibitem [{\citenamefont {Mari}\ \emph {et~al.}(2014)\citenamefont {Mari},
  \citenamefont {Seto}, \citenamefont {Morris},\ and\ \citenamefont
  {Denn}}]{Mari_2014JOR}%
  \BibitemOpen
  \bibfield  {author} {\bibinfo {author} {\bibfnamefont {R.}~\bibnamefont
  {Mari}}, \bibinfo {author} {\bibfnamefont {R.}~\bibnamefont {Seto}}, \bibinfo
  {author} {\bibfnamefont {J.~F.}\ \bibnamefont {Morris}},\ and\ \bibinfo
  {author} {\bibfnamefont {M.~M.}\ \bibnamefont {Denn}},\ }\href
  {https://doi.org/10.1122/1.4890747} {\bibfield  {journal} {\bibinfo
  {journal} {J. Rheol.}\ }\textbf {\bibinfo {volume} {58}},\ \bibinfo {pages}
  {1693} (\bibinfo {year} {2014})},\ \Eprint
  {https://arxiv.org/abs/https://doi.org/10.1122/1.4890747}
  {https://doi.org/10.1122/1.4890747} \BibitemShut {NoStop}%
\bibitem [{Note2()}]{Note2}%
  \BibitemOpen
  \bibinfo {note} {The Cox-Merz rule states that the dynamic viscosity of a
  fluid in SS is equal to its complex viscosity in OS at matching frequency,
  \protect \textit {i.e.~}$\eta (\protect \mathaccentV {dot}05F{\gamma })=\eta
  ^*(\omega )|_{\omega =\protect \mathaccentV {dot}05F{\gamma }}$; see Ref.\
  \cite {mewis_wagner_book}.}\BibitemShut {Stop}%
\bibitem [{\citenamefont {Bricker}\ and\ \citenamefont
  {Butler}(2006)}]{Bricker_Butler2006}%
  \BibitemOpen
  \bibfield  {author} {\bibinfo {author} {\bibfnamefont {J.~M.}\ \bibnamefont
  {Bricker}}\ and\ \bibinfo {author} {\bibfnamefont {J.~E.}\ \bibnamefont
  {Butler}},\ }\href {https://doi.org/10.1122/1.2234366} {\bibfield  {journal}
  {\bibinfo  {journal} {J. Rheol.}\ }\textbf {\bibinfo {volume} {50}},\
  \bibinfo {pages} {711} (\bibinfo {year} {2006})},\ \Eprint
  {https://arxiv.org/abs/https://doi.org/10.1122/1.2234366}
  {https://doi.org/10.1122/1.2234366} \BibitemShut {NoStop}%
\bibitem [{\citenamefont {Bricker}\ and\ \citenamefont
  {Butler}(2007)}]{Bricker_Butler2007}%
  \BibitemOpen
  \bibfield  {author} {\bibinfo {author} {\bibfnamefont {J.~M.}\ \bibnamefont
  {Bricker}}\ and\ \bibinfo {author} {\bibfnamefont {J.~E.}\ \bibnamefont
  {Butler}},\ }\href {https://doi.org/10.1122/1.2724886} {\bibfield  {journal}
  {\bibinfo  {journal} {Journal of Rheology}\ }\textbf {\bibinfo {volume}
  {51}},\ \bibinfo {pages} {735} (\bibinfo {year} {2007})},\ \Eprint
  {https://arxiv.org/abs/https://doi.org/10.1122/1.2724886}
  {https://doi.org/10.1122/1.2724886} \BibitemShut {NoStop}%
\bibitem [{\citenamefont {Carotenuto}\ \emph {et~al.}(2014)\citenamefont
  {Carotenuto}, \citenamefont {Merola},\ and\ \citenamefont
  {Minale}}]{Carotenuto_conf_2014}%
  \BibitemOpen
  \bibfield  {author} {\bibinfo {author} {\bibfnamefont {C.}~\bibnamefont
  {Carotenuto}}, \bibinfo {author} {\bibfnamefont {M.~C.}\ \bibnamefont
  {Merola}},\ and\ \bibinfo {author} {\bibfnamefont {M.}~\bibnamefont
  {Minale}},\ }\href {https://doi.org/10.1063/1.4876827} {\bibfield  {journal}
  {\bibinfo  {journal} {AIP Conference Proceedings}\ }\textbf {\bibinfo
  {volume} {1599}},\ \bibinfo {pages} {258} (\bibinfo {year} {2014})},\ \Eprint
  {https://arxiv.org/abs/https://aip.scitation.org/doi/pdf/10.1063/1.4876827}
  {https://aip.scitation.org/doi/pdf/10.1063/1.4876827} \BibitemShut {NoStop}%
\bibitem [{\citenamefont {Martone}\ \emph {et~al.}(2018)\citenamefont
  {Martone}, \citenamefont {Paduano}, \citenamefont {Carotenuto},\ and\
  \citenamefont {Minale}}]{Martone_conf_2018}%
  \BibitemOpen
  \bibfield  {author} {\bibinfo {author} {\bibfnamefont {R.}~\bibnamefont
  {Martone}}, \bibinfo {author} {\bibfnamefont {L.~P.}\ \bibnamefont
  {Paduano}}, \bibinfo {author} {\bibfnamefont {C.}~\bibnamefont
  {Carotenuto}},\ and\ \bibinfo {author} {\bibfnamefont {M.}~\bibnamefont
  {Minale}},\ }\href {https://doi.org/10.1063/1.5046047} {\bibfield  {journal}
  {\bibinfo  {journal} {AIP Conference Proceedings}\ }\textbf {\bibinfo
  {volume} {1981}},\ \bibinfo {pages} {020185} (\bibinfo {year} {2018})},\
  \Eprint
  {https://arxiv.org/abs/https://aip.scitation.org/doi/pdf/10.1063/1.5046047}
  {https://aip.scitation.org/doi/pdf/10.1063/1.5046047} \BibitemShut {NoStop}%
\bibitem [{\citenamefont {Martone}\ \emph {et~al.}(2020)\citenamefont
  {Martone}, \citenamefont {Carotenuto},\ and\ \citenamefont
  {Minale}}]{Martone_experiment}%
  \BibitemOpen
  \bibfield  {author} {\bibinfo {author} {\bibfnamefont {R.}~\bibnamefont
  {Martone}}, \bibinfo {author} {\bibfnamefont {C.}~\bibnamefont
  {Carotenuto}},\ and\ \bibinfo {author} {\bibfnamefont {M.}~\bibnamefont
  {Minale}},\ }\href@noop {} {\bibfield  {journal} {\bibinfo  {journal}
  {submitted to J. Rheol.}\ } (\bibinfo {year} {2020})}\BibitemShut {NoStop}%
\bibitem [{\citenamefont {Del~Gado}\ and\ \citenamefont
  {Morris}(2020)}]{dense_preface2020}%
  \BibitemOpen
  \bibfield  {author} {\bibinfo {author} {\bibfnamefont {E.}~\bibnamefont
  {Del~Gado}}\ and\ \bibinfo {author} {\bibfnamefont {J.~F.}\ \bibnamefont
  {Morris}},\ }\href {https://doi.org/10.1122/8.0000016} {\bibfield  {journal}
  {\bibinfo  {journal} {Journal of Rheology}\ }\textbf {\bibinfo {volume}
  {64}},\ \bibinfo {pages} {223} (\bibinfo {year} {2020})},\ \Eprint
  {https://arxiv.org/abs/https://doi.org/10.1122/8.0000016}
  {https://doi.org/10.1122/8.0000016} \BibitemShut {NoStop}%
\bibitem [{\citenamefont {Breedveld}\ \emph {et~al.}(2001)\citenamefont
  {Breedveld}, \citenamefont {van~den Ende}, \citenamefont {Jongschaap},\ and\
  \citenamefont {Mellema}}]{Breedveld_etal_2001}%
  \BibitemOpen
  \bibfield  {author} {\bibinfo {author} {\bibfnamefont {V.}~\bibnamefont
  {Breedveld}}, \bibinfo {author} {\bibfnamefont {D.}~\bibnamefont {van~den
  Ende}}, \bibinfo {author} {\bibfnamefont {R.}~\bibnamefont {Jongschaap}},\
  and\ \bibinfo {author} {\bibfnamefont {J.}~\bibnamefont {Mellema}},\ }\href
  {https://doi.org/10.1063/1.1355315} {\bibfield  {journal} {\bibinfo
  {journal} {J. Chem. Phys.}\ }\textbf {\bibinfo {volume} {114}},\ \bibinfo
  {pages} {5923} (\bibinfo {year} {2001})},\ \Eprint
  {https://arxiv.org/abs/https://doi.org/10.1063/1.1355315}
  {https://doi.org/10.1063/1.1355315} \BibitemShut {NoStop}%
\bibitem [{\citenamefont {Cort\'e}\ \emph {et~al.}(2008)\citenamefont
  {Cort\'e}, \citenamefont {Chaikin}, \citenamefont {Gollub},\ and\
  \citenamefont {Pine}}]{Corte_NatPhys_2008}%
  \BibitemOpen
  \bibfield  {author} {\bibinfo {author} {\bibfnamefont {L.}~\bibnamefont
  {Cort\'e}}, \bibinfo {author} {\bibfnamefont {P.~M.}\ \bibnamefont
  {Chaikin}}, \bibinfo {author} {\bibfnamefont {J.~P.}\ \bibnamefont
  {Gollub}},\ and\ \bibinfo {author} {\bibfnamefont {D.~J.}\ \bibnamefont
  {Pine}},\ }\href@noop {} {\bibfield  {journal} {\bibinfo  {journal} {Nat.
  Phys.}\ }\textbf {\bibinfo {volume} {4}},\ \bibinfo {pages} {420} (\bibinfo
  {year} {2008})}\BibitemShut {NoStop}%
\bibitem [{\citenamefont {Lin}\ \emph {et~al.}(2013)\citenamefont {Lin},
  \citenamefont {Phan-Thien},\ and\ \citenamefont
  {Khoo}}]{Lin_Phan-Thien_Khoo_2013}%
  \BibitemOpen
  \bibfield  {author} {\bibinfo {author} {\bibfnamefont {Y.}~\bibnamefont
  {Lin}}, \bibinfo {author} {\bibfnamefont {N.}~\bibnamefont {Phan-Thien}},\
  and\ \bibinfo {author} {\bibfnamefont {B.~C.}\ \bibnamefont {Khoo}},\ }\href
  {https://doi.org/10.1122/1.4815979} {\bibfield  {journal} {\bibinfo
  {journal} {J. Rheol.}\ }\textbf {\bibinfo {volume} {57}},\ \bibinfo {pages}
  {1325} (\bibinfo {year} {2013})},\ \Eprint
  {https://arxiv.org/abs/https://doi.org/10.1122/1.4815979}
  {https://doi.org/10.1122/1.4815979} \BibitemShut {NoStop}%
\bibitem [{\citenamefont {Cheal}\ and\ \citenamefont
  {Ness}(2018)}]{Cheal_Ness_2018}%
  \BibitemOpen
  \bibfield  {author} {\bibinfo {author} {\bibfnamefont {O.}~\bibnamefont
  {Cheal}}\ and\ \bibinfo {author} {\bibfnamefont {C.}~\bibnamefont {Ness}},\
  }\href {https://doi.org/10.1122/1.5004007} {\bibfield  {journal} {\bibinfo
  {journal} {Journal of Rheology}\ }\textbf {\bibinfo {volume} {62}},\ \bibinfo
  {pages} {501} (\bibinfo {year} {2018})},\ \Eprint
  {https://arxiv.org/abs/https://doi.org/10.1122/1.5004007}
  {https://doi.org/10.1122/1.5004007} \BibitemShut {NoStop}%
\bibitem [{\citenamefont {Ge}\ and\ \citenamefont {Brandt}(2020)}]{hlgd}%
  \BibitemOpen
  \bibfield  {author} {\bibinfo {author} {\bibfnamefont {Z.}~\bibnamefont
  {Ge}}\ and\ \bibinfo {author} {\bibfnamefont {L.}~\bibnamefont {Brandt}},\
  }\href@noop {} {\bibinfo {title} {Implementation note on a minimal hybrid
  lubrication/granular dynamics model for dense suspensions}} (\bibinfo {year}
  {2020}),\ \bibinfo {note} {arXiv:2005.12755}\BibitemShut {NoStop}%
\bibitem [{Note3()}]{Note3}%
  \BibitemOpen
  \bibinfo {note} {There is yet another phenomenological mechanism for shear
  thinning -- stress relaxation. Since the sample is presheared in SS then
  measured in OS, at $\gamma _0 \lessapprox 1$, the previously accumulated
  stress may be released to a greater extent in lower $\gamma _0$. However,
  such relaxation has no time dependence (no additional force scale), thus
  cannot explain the present experiment.}\BibitemShut {Stop}%
\bibitem [{\citenamefont {V\'azquez-Quesada}\ \emph {et~al.}(2016)\citenamefont
  {V\'azquez-Quesada}, \citenamefont {Tanner},\ and\ \citenamefont
  {Ellero}}]{NN_solvent0}%
  \BibitemOpen
  \bibfield  {author} {\bibinfo {author} {\bibfnamefont {A.}~\bibnamefont
  {V\'azquez-Quesada}}, \bibinfo {author} {\bibfnamefont {R.~I.}\ \bibnamefont
  {Tanner}},\ and\ \bibinfo {author} {\bibfnamefont {M.}~\bibnamefont
  {Ellero}},\ }\href {https://doi.org/10.1103/PhysRevLett.117.108001}
  {\bibfield  {journal} {\bibinfo  {journal} {Phys. Rev. Lett.}\ }\textbf
  {\bibinfo {volume} {117}},\ \bibinfo {pages} {108001} (\bibinfo {year}
  {2016})}\BibitemShut {NoStop}%
\bibitem [{\citenamefont {V\'azquez-Quesada}\ \emph {et~al.}(2017)\citenamefont
  {V\'azquez-Quesada}, \citenamefont {Mahmud}, \citenamefont {Dai},
  \citenamefont {Ellero},\ and\ \citenamefont {Tanner}}]{NN_solvent1}%
  \BibitemOpen
  \bibfield  {author} {\bibinfo {author} {\bibfnamefont {A.}~\bibnamefont
  {V\'azquez-Quesada}}, \bibinfo {author} {\bibfnamefont {A.}~\bibnamefont
  {Mahmud}}, \bibinfo {author} {\bibfnamefont {S.}~\bibnamefont {Dai}},
  \bibinfo {author} {\bibfnamefont {M.}~\bibnamefont {Ellero}},\ and\ \bibinfo
  {author} {\bibfnamefont {R.~I.}\ \bibnamefont {Tanner}},\ }\href
  {https://doi.org/https://doi.org/10.1016/j.jnnfm.2017.08.005} {\bibfield
  {journal} {\bibinfo  {journal} {Journal of Non-Newtonian Fluid Mechanics}\
  }\textbf {\bibinfo {volume} {248}},\ \bibinfo {pages} {1 } (\bibinfo {year}
  {2017})}\BibitemShut {NoStop}%
\bibitem [{\citenamefont {Kalman}\ \emph {et~al.}(2008)\citenamefont {Kalman},
  \citenamefont {Rosen},\ and\ \citenamefont
  {Wagner}}]{Kalman_Rosen_Wagner2008}%
  \BibitemOpen
  \bibfield  {author} {\bibinfo {author} {\bibfnamefont {D.~P.}\ \bibnamefont
  {Kalman}}, \bibinfo {author} {\bibfnamefont {B.~A.}\ \bibnamefont {Rosen}},\
  and\ \bibinfo {author} {\bibfnamefont {N.~J.}\ \bibnamefont {Wagner}},\
  }\href {https://doi.org/10.1063/1.2964591} {\bibfield  {journal} {\bibinfo
  {journal} {AIP Conference Proceedings}\ }\textbf {\bibinfo {volume} {1027}},\
  \bibinfo {pages} {1408} (\bibinfo {year} {2008})},\ \Eprint
  {https://arxiv.org/abs/https://aip.scitation.org/doi/pdf/10.1063/1.2964591}
  {https://aip.scitation.org/doi/pdf/10.1063/1.2964591} \BibitemShut {NoStop}%
\bibitem [{\citenamefont {Radhakrishnan}\ and\ \citenamefont
  {Sun}(2019)}]{ranga_thinning}%
  \BibitemOpen
  \bibfield  {author} {\bibinfo {author} {\bibfnamefont {R.}~\bibnamefont
  {Radhakrishnan}}\ and\ \bibinfo {author} {\bibfnamefont {J.}~\bibnamefont
  {Sun}}} (\bibinfo {year} {2019}),\ \bibinfo {note} {unpublished}\BibitemShut
  {NoStop}%
\bibitem [{\citenamefont {Irani}\ \emph {et~al.}(2019)\citenamefont {Irani},
  \citenamefont {Chaudhuri},\ and\ \citenamefont {Heussinger}}]{Irani_prf2019}%
  \BibitemOpen
  \bibfield  {author} {\bibinfo {author} {\bibfnamefont {E.}~\bibnamefont
  {Irani}}, \bibinfo {author} {\bibfnamefont {P.}~\bibnamefont {Chaudhuri}},\
  and\ \bibinfo {author} {\bibfnamefont {C.}~\bibnamefont {Heussinger}},\
  }\href {https://doi.org/10.1103/PhysRevFluids.4.074307} {\bibfield  {journal}
  {\bibinfo  {journal} {Phys. Rev. Fluids}\ }\textbf {\bibinfo {volume} {4}},\
  \bibinfo {pages} {074307} (\bibinfo {year} {2019})}\BibitemShut {NoStop}%
\bibitem [{\citenamefont {Lobry}\ \emph {et~al.}(2019)\citenamefont {Lobry},
  \citenamefont {Lemaire}, \citenamefont {Blanc}, \citenamefont {Gallier},\
  and\ \citenamefont {Peters}}]{lobry_lemaire_blanc_gallier_peters_2019}%
  \BibitemOpen
  \bibfield  {author} {\bibinfo {author} {\bibfnamefont {L.}~\bibnamefont
  {Lobry}}, \bibinfo {author} {\bibfnamefont {E.}~\bibnamefont {Lemaire}},
  \bibinfo {author} {\bibfnamefont {F.}~\bibnamefont {Blanc}}, \bibinfo
  {author} {\bibfnamefont {S.}~\bibnamefont {Gallier}},\ and\ \bibinfo {author}
  {\bibfnamefont {F.}~\bibnamefont {Peters}},\ }\href
  {https://doi.org/10.1017/jfm.2018.881} {\bibfield  {journal} {\bibinfo
  {journal} {Journal of Fluid Mechanics}\ }\textbf {\bibinfo {volume} {860}},\
  \bibinfo {pages} {682–710} (\bibinfo {year} {2019})}\BibitemShut {NoStop}%
\bibitem [{\citenamefont {Brown}\ \emph {et~al.}(2010)\citenamefont {Brown},
  \citenamefont {Forman}, \citenamefont {Orellana}, \citenamefont {Zhang},
  \citenamefont {Maynor}, \citenamefont {Betts}, \citenamefont {DeSimone},\
  and\ \citenamefont {Jaeger}}]{Brown_nmat2010}%
  \BibitemOpen
  \bibfield  {author} {\bibinfo {author} {\bibfnamefont {E.}~\bibnamefont
  {Brown}}, \bibinfo {author} {\bibfnamefont {N.~A.}\ \bibnamefont {Forman}},
  \bibinfo {author} {\bibfnamefont {C.~S.}\ \bibnamefont {Orellana}}, \bibinfo
  {author} {\bibfnamefont {H.}~\bibnamefont {Zhang}}, \bibinfo {author}
  {\bibfnamefont {B.~W.}\ \bibnamefont {Maynor}}, \bibinfo {author}
  {\bibfnamefont {D.~E.}\ \bibnamefont {Betts}}, \bibinfo {author}
  {\bibfnamefont {J.~M.}\ \bibnamefont {DeSimone}},\ and\ \bibinfo {author}
  {\bibfnamefont {H.~M.}\ \bibnamefont {Jaeger}},\ }\href
  {https://doi.org/10.1038/nmat2627} {\bibfield  {journal} {\bibinfo  {journal}
  {Nature Materials}\ }\textbf {\bibinfo {volume} {9}},\ \bibinfo {pages} {220}
  (\bibinfo {year} {2010})}\BibitemShut {NoStop}%
\bibitem [{\citenamefont {Singh}\ \emph {et~al.}(2019)\citenamefont {Singh},
  \citenamefont {Pednekar}, \citenamefont {Chun}, \citenamefont {Denn},\ and\
  \citenamefont {Morris}}]{Singh_attr_prl2019}%
  \BibitemOpen
  \bibfield  {author} {\bibinfo {author} {\bibfnamefont {A.}~\bibnamefont
  {Singh}}, \bibinfo {author} {\bibfnamefont {S.}~\bibnamefont {Pednekar}},
  \bibinfo {author} {\bibfnamefont {J.}~\bibnamefont {Chun}}, \bibinfo {author}
  {\bibfnamefont {M.~M.}\ \bibnamefont {Denn}},\ and\ \bibinfo {author}
  {\bibfnamefont {J.~F.}\ \bibnamefont {Morris}},\ }\href
  {https://doi.org/10.1103/PhysRevLett.122.098004} {\bibfield  {journal}
  {\bibinfo  {journal} {Phys. Rev. Lett.}\ }\textbf {\bibinfo {volume} {122}},\
  \bibinfo {pages} {098004} (\bibinfo {year} {2019})}\BibitemShut {NoStop}%
\bibitem [{Note4()}]{Note4}%
  \BibitemOpen
  \bibinfo {note} {The maximal shear stress in the experiment is $\sigma
  _s=1120$ Pa, way below the elastic modulus of the glass microspheres ($G
  \approx 70$ GPa). Thus, the deformation is negligible ($\sigma _s/G <
  1.6\times 10^{-8}$).}\BibitemShut {Stop}%
\bibitem [{\citenamefont {Israelachvili}(2011)}]{Israelachvili_book}%
  \BibitemOpen
  \bibfield  {author} {\bibinfo {author} {\bibfnamefont {J.}~\bibnamefont
  {Israelachvili}},\ }\href@noop {} {\emph {\bibinfo {title} {Intermolecular
  and Surface Forces}}}\ (\bibinfo  {publisher} {Academic Press},\ \bibinfo
  {year} {2011})\BibitemShut {NoStop}%
\bibitem [{Note5()}]{Note5}%
  \BibitemOpen
  \bibinfo {note} {See Fig.\ 10 of Ref.\ \cite {Martone_experiment} for
  evidence of the $\phi $ dependence.}\BibitemShut {Stop}%
\bibitem [{\citenamefont {Poole}(2012)}]{Poole_Wi_De}%
  \BibitemOpen
  \bibfield  {author} {\bibinfo {author} {\bibfnamefont {R.}~\bibnamefont
  {Poole}},\ }\href@noop {} {\bibfield  {journal} {\bibinfo  {journal} {The
  British Society of Rheology, Rheology Bulletin.}\ }\textbf {\bibinfo {volume}
  {53(2)}},\ \bibinfo {pages} {32} (\bibinfo {year} {2012})}\BibitemShut
  {NoStop}%
\bibitem [{\citenamefont {Ness}\ \emph {et~al.}(2017)\citenamefont {Ness},
  \citenamefont {Xing},\ and\ \citenamefont {Eiser}}]{Ness_OS_2017SM}%
  \BibitemOpen
  \bibfield  {author} {\bibinfo {author} {\bibfnamefont {C.}~\bibnamefont
  {Ness}}, \bibinfo {author} {\bibfnamefont {Z.}~\bibnamefont {Xing}},\ and\
  \bibinfo {author} {\bibfnamefont {E.}~\bibnamefont {Eiser}},\ }\href
  {https://doi.org/10.1039/C7SM00039A} {\bibfield  {journal} {\bibinfo
  {journal} {Soft Matter}\ }\textbf {\bibinfo {volume} {13}},\ \bibinfo {pages}
  {3664} (\bibinfo {year} {2017})}\BibitemShut {NoStop}%
\bibitem [{Note6()}]{Note6}%
  \BibitemOpen
  \bibinfo {note} {An alternative measure of the microstructure evolution is
  particle diffusivities, cf.\ \cite {Pine_Nature_2005}. We have checked the
  cases of $\gamma _0=0.1$ and $\omega =0.1$ under different Sr$_{attr}$, and
  found negligible self-diffusivities throughout.}\BibitemShut {Stop}%
\bibitem [{\citenamefont {Wagner}\ and\ \citenamefont
  {Brady}(2009)}]{Wagner_Brady_2009}%
  \BibitemOpen
  \bibfield  {author} {\bibinfo {author} {\bibfnamefont {N.~J.}\ \bibnamefont
  {Wagner}}\ and\ \bibinfo {author} {\bibfnamefont {J.~F.}\ \bibnamefont
  {Brady}},\ }\href@noop {} {\bibfield  {journal} {\bibinfo  {journal} {Physics
  Today}\ }\textbf {\bibinfo {volume} {62 (10)}},\ \bibinfo {pages} {27}
  (\bibinfo {year} {2009})}\BibitemShut {NoStop}%
\bibitem [{\citenamefont {Cort\'e}\ \emph {et~al.}(2009)\citenamefont
  {Cort\'e}, \citenamefont {Gerbode}, \citenamefont {Man},\ and\ \citenamefont
  {Pine}}]{Corte_etal_PRL2009}%
  \BibitemOpen
  \bibfield  {author} {\bibinfo {author} {\bibfnamefont {L.}~\bibnamefont
  {Cort\'e}}, \bibinfo {author} {\bibfnamefont {S.~J.}\ \bibnamefont
  {Gerbode}}, \bibinfo {author} {\bibfnamefont {W.}~\bibnamefont {Man}},\ and\
  \bibinfo {author} {\bibfnamefont {D.~J.}\ \bibnamefont {Pine}},\ }\href
  {https://doi.org/10.1103/PhysRevLett.103.248301} {\bibfield  {journal}
  {\bibinfo  {journal} {Phys. Rev. Lett.}\ }\textbf {\bibinfo {volume} {103}},\
  \bibinfo {pages} {248301} (\bibinfo {year} {2009})}\BibitemShut {NoStop}%
\bibitem [{\citenamefont {Royer}\ and\ \citenamefont
  {Chaikin}(2015)}]{Royer49}%
  \BibitemOpen
  \bibfield  {author} {\bibinfo {author} {\bibfnamefont {J.~R.}\ \bibnamefont
  {Royer}}\ and\ \bibinfo {author} {\bibfnamefont {P.~M.}\ \bibnamefont
  {Chaikin}},\ }\href {https://doi.org/10.1073/pnas.1413468112} {\bibfield
  {journal} {\bibinfo  {journal} {Proceedings of the National Academy of
  Sciences}\ }\textbf {\bibinfo {volume} {112}},\ \bibinfo {pages} {49}
  (\bibinfo {year} {2015})},\ \Eprint
  {https://arxiv.org/abs/https://www.pnas.org/content/112/1/49.full.pdf}
  {https://www.pnas.org/content/112/1/49.full.pdf} \BibitemShut {NoStop}%
\bibitem [{\citenamefont {Ness}\ and\ \citenamefont
  {Cates}(2020)}]{Ness_Cates_PRL2020}%
  \BibitemOpen
  \bibfield  {author} {\bibinfo {author} {\bibfnamefont {C.}~\bibnamefont
  {Ness}}\ and\ \bibinfo {author} {\bibfnamefont {M.~E.}\ \bibnamefont
  {Cates}},\ }\href {https://doi.org/10.1103/PhysRevLett.124.088004} {\bibfield
   {journal} {\bibinfo  {journal} {Phys. Rev. Lett.}\ }\textbf {\bibinfo
  {volume} {124}},\ \bibinfo {pages} {088004} (\bibinfo {year}
  {2020})}\BibitemShut {NoStop}%
\bibitem [{\citenamefont {Pine}\ \emph {et~al.}(2005)\citenamefont {Pine},
  \citenamefont {Gollub}, \citenamefont {Brady},\ and\ \citenamefont
  {Leshansky}}]{Pine_Nature_2005}%
  \BibitemOpen
  \bibfield  {author} {\bibinfo {author} {\bibfnamefont {D.~J.}\ \bibnamefont
  {Pine}}, \bibinfo {author} {\bibfnamefont {J.~P.}\ \bibnamefont {Gollub}},
  \bibinfo {author} {\bibfnamefont {J.~F.}\ \bibnamefont {Brady}},\ and\
  \bibinfo {author} {\bibfnamefont {A.~M.}\ \bibnamefont {Leshansky}},\
  }\href@noop {} {\bibfield  {journal} {\bibinfo  {journal} {Nature}\ }\textbf
  {\bibinfo {volume} {438}},\ \bibinfo {pages} {1476} (\bibinfo {year}
  {2005})}\BibitemShut {NoStop}%
\end{thebibliography}%



\end{document}